\documentclass[10pt]{article}

\usepackage{amssymb}
\usepackage{amsmath}
\usepackage{amsthm} 
\usepackage{mathrsfs}                                                           
\usepackage{slashbox}  
\usepackage[all]{xy}
\usepackage{graphicx}
\usepackage{wrapfig}

\def\2{{1\over 2}}

\newcommand{\rf}[1]{(\ref{#1})}
\def\b{\bar}

\newcommand{\ud}{\mathrm{d}}
\renewcommand{\t}{\tilde}
\newcommand{\p}{\partial}
\newcommand{\bp}{\bar{\partial}}

\newcommand{\mA}{\mathbf{A} }
\newcommand{\mB}{\mathbf{B} }

\newcommand{\mV}{\mathbf{V} }
\newcommand{\mW}{\mathbf{W} }

\newcommand{\mg}{\mathfrak{g}}

\title{\bf{$\beta$-$\gamma$ systems and the deformations of the BRST operator}}
\author{Anton M. Zeitlin\footnote{anton.zeitlin@yale.edu, http://math.yale.edu/$\sim$az84, http://www.ipme.ru/zam.html} \\
Department of Mathematics,\\
Yale University,\\
442 Dunham Lab, 10 Hillhouse Ave\\
New Haven, CT 06511}
\date{}

\begin{document}
\maketitle
\begin{abstract}
We describe the relation between simple logarithmic CFTs associated with closed and open strings, and their "infinite metric" limits, corresponding to the $\beta$-$\gamma$ systems. This relation is studied on the level of the BRST complex: we show that the consideration of metric as a perturbation leads to a certain deformation of the algebraic operations of the Lian-Zuckerman type on the vertex algebra, associated with the $\beta$-$\gamma$ systems. The Maurer-Cartan equations corresponding to this deformed structure in the quasiclassical approximation lead to the nonlinear field equations. As an explicit example, we demonstrate, that using this construction, Yang-Mills equations can be derived. This gives rise to a nontrivial relation between the Courant-Dorfman algebroid and homotopy algebras emerging from the gauge theory. We also discuss possible algebraic approach to the study of beta-functions in sigma-models.
\end{abstract}
\section{Introduction}
It is well-known that string theory is a tool, that allows to derive the various properties of the "target space" (which is $D$-dimensional) from the two dimensional "worldsheet". The theory of the so-called vertex operator algebras (see e.g. \cite{fhl}) serves as a mathematical method in the investigation of the two dimensional (quantum) world. They also turn out to be helpful in the relation  between the worldsheet and target space. It is worth mentioning recent papers on the chiral de Rham complex, sheaves of vertex algebras and their applications to pure spinor superstrings  and  topological strings  (see e.g. \cite{malikov}-\cite{nekrasov}). 

One can ask a question: what is the meaning of the classical field equations from a vertex operator algebra perspective? In this article, following the considerations of \cite{cftym}, we try to solve this conundrum.  

It is known (see e.g. \cite{bvym}) that the nonlinear equations of the field theory such as the Yang-Mills (YM) equations originate as Maurer-Cartan equations for certain homotopy algebras. The String Field Theory (SFT) \cite{witten}, \cite{zwiebach}, \cite{stakaj} suggests that these algebras should be 
reconstructed from some operations on the two-dimensional worldsheet. However, it is extremely hard to derive them directly from SFT, since one has to integrate out massive modes (see e.g. \cite{taylor}, \cite{berkovits}). 

We show, that there is another way, through the homotopy algebras, related to the vertex operator algebra, which were introduced by Lian and Zuckerman \cite{lz}. One can associate a semi-infinite (BRST) complex to any vertex operator 
algebra (with the central charge of the Virasoro algebra equal to 26 \cite{fgz}) by  extending it via the conformal $b$-$c$ ghost system. This complex has a natural multiplication, which is homotopy associative and commutative, and another bilinear operation with ghost number equal to $-1$. Together, they satisfy the relations of the homotopy Gerstenhaber algebra. Lian-Zuckerman algebras already proved to be useful: they determined the algebraic structure on the ground rings of 2d Gravity \cite{wz} (see also \cite{fz}). But in this case the main interest was in the corresponding BRST cohomology algebra, and the homotopical nature of the algebra was totally  neglected.  

In \cite{cftym} we have shown, that applying the Lian-Zuckerman (LZ) operations directly to the logarithmic operator algebra of the open string (neglecting logarithms) one is able to reproduce the $C_{\infty}$ algebra of the Yang-Mills theory on the quasiclassical level (neglecting higher order $\alpha'$-corrections). However, it seems to be a miraculous coincidence, since the LZ approach does not work for logarithmic vertex algebras (see e.g. \cite{flohr}, \cite{semikhatov}).  Here we demonstrate, that actually it is not a coincidence. We show how to embed the BRST complex of the open string into the deformation of the BRST complex, associated with the corresponding $\beta$-$\gamma$ system. 

We explicitly show that the resulting deformed LZ structure reproduces the YM equations and their symmetries, explaining the results of \cite{cftym} on the vertex operator algebra level. One of the immediate important consequences of this approach is a relation between the YM $C_{\infty}$ algebra and the Courant-Dorfman brackets, which naturally appear in the study of $\beta$-$\gamma$ systems (see e.g. \cite{zeit2}).

The initial idea about reproducing classical equations 
allows the following extension. It is well known, that the beta-function of nonlinear sigma-models reproduces classical equations at the first order in $\alpha'$. However, the whole expression for the beta-function (all $\alpha'$-corrections) is not yet known. Moreover, it is not unique, because it depends on the regularization scheme. The original idea of algebraization of beta-functions in 2d from the  perspective of the deformation of the BRST operator goes back to classical papers from 80s (see e.g. \cite{banks}). This is, however, precisely what the Lian-Zuckerman operations do. If one extends further the corresponding homotopy algebra to the so-called $G_{\infty}$-algebra  \cite{zuck}, \cite{gorbounov}, one can interpret the right hand side of the corresponding Maurer-Cartan equation as an algebraic definition of the beta-function. This  gives a promising approach to derive all the  $\alpha'$-corrections. 

The article is organized as follows. In Section 2, we recall basic facts about the LZ  homotopy Gerstenhaber algebra and explain its relation to the perturbed 2d CFTs. In Section 3, we use these constructions in order to obtain the BRST operator in logarithmic theories associated with open and closed strings as a deformation of the BRST operator for the certain ($\beta$-$\gamma$) first order theory, 
which can be described without logarithms i.e. via the vertex operator algebra. In particular, for the open string this leads to big advantages: it is possible to isolate 
light modes as a BRST subcomplex and to apply the Lian-Zuckerman construction to this subcomplex only, neglecting massive modes. 

In Section 4, we study a semiclassical approximation to the Maurer-Cartan equation associated with the  homotopy associative algebra (which is a subalgebra of the LZ  homotopy Gerstenhaber algebra). It appears that in the case of the open string reformulated via the $\beta$-$\gamma$ system it leads to the YM equation, confirming the "surprising" results of \cite{cftym}, where we applied the LZ operations directly to the case of the logarithmic CFT of the open string.  In Section 5, we outline further developments.

\section{Deformation of charges via the Lian-Zuckerman operations}
In this section, we give necessary facts about the Lian-Zuckerman operations \cite{lz}, although from a different insight, which we need in our approach. \\

\noindent {\bf 2.1. Reminder of the Lian-Zuckerman homotopy algebras.} Let us consider some chiral algebra $V$. Let $T(z)$ denote the appropriate Virasoro element. Let $\Lambda^*$ denote the space of states of the conformal $b$-$c$ ghost system. Then  one 
 can define an operator $Q$ acting on the $semi$-$infinite$ complex $C^*=V\otimes \Lambda^*$:
\begin{eqnarray}\label{brst}
Q=\oint dz (c(z)T(z)+:bc\p c(z):). 
\end{eqnarray}
This operator is known as the BRST operator associated with the chiral algebra 
$V$. It is well-known that $Q$ is nilpotent on $C^*$ when the central charge of the Virasoro algebra associated with $T(z)$ is equal to 26.  

Let $a(z)$ be a vertex operator associated with the state $a$. Then one can define the following bilinear operation on the corresponding space of states:
\begin{eqnarray}\label{mu}
\mu(a_1,a_2)=P_0a_1(\epsilon)a_2,
\end{eqnarray}
where  $P_0$ is the projection on $\epsilon$-independent part, (the right hand side is considered as a power series in $\epsilon$). It was shown \cite{lz} that this bilinear operation is homotopy commutative and associative w.r.t. the operator $Q$, namely  
the bilinear operation $\mu$ satisfies the following relations:
\begin{eqnarray}\label{lzrel}
&&Q\mu(a_1,a_2)=\mu(Q a_1,a_2)+(-1)^{|a_1|}\mu(a_1,Q a_2),\nonumber\\
&&\mu(a_1,a_2)-(-1)^{|a_1||a_2|}\mu(a_2,a_1)=\nonumber\\
&&Qm(a_1,a_2)+m(Qa_1,a_2)+(-1)^{|a_1|}m(a_1,Qa_2),\nonumber\\
&&\mu(\mu(a_1,a_2),a_3)-\mu(a_1,\mu(a_2,a_3))=\nonumber\\
&&Qn(a_1,a_2,a_3)+n(Qa_1,a_2,a_3)+(-1)^{|a_1|}n(a_1,Qa_2,a_3)+\nonumber\\
&&(-1)^{|a_1|+|a_2|}n(a_1,a_2,Qa_3),
\end{eqnarray}
where 
\begin{eqnarray}
&&m(a_1,a_2)=\sum_{i\ge 0}\frac{(-1)^i}{i+1}Res_wRes_{z-w}(z-w)^iw^{-i-1}b_{-1}\nonumber\\
&&\qquad\qquad\quad (a_1(z-w)a_2)(w)\mathbf{1},
\nonumber\\
&&n(a_1,a_2,a_3)=\sum_{i\ge 0}\frac{1}{i+1}Res_zRes_w w^iz^{-i-1}(b_{-1}a_1)(z)a_2(w)a_3+\nonumber\\
&&(-1)^{|a_1||a_2|}\sum_{i\ge 0}\frac{1}{i+1}Res_wRes_z z^iw^{-i-1}(b_{-1}a_2)(w)a_1(z)a_3.
\end{eqnarray}
Thus, when $Q$ is nilpotent, on the level of cohomology w.r.t. the operator $Q$, this algebra turns out to be commutative and associative. However, originally on the BRST complex, this operation is only {\it homotopy associative}. 

One is able to define another useful bilinear operation:
\begin{eqnarray}\label{gerst}
\{a_1,a_2\}=\frac{(-1)^{|a_1|}}{2\pi i} \oint dz (b_{-1}a_1)(z)a_2
\end{eqnarray}
which, together with operation $\mu$, satisfies the relations of the homotopy  Gerstenhaber algebra: 
\begin{eqnarray}
&&\{a_1,a_2\}+(-1)^{(|a_1|-1)(|a_2|-1)}\{a_2,a_1\}=\\
&&(-1)^{|a_1|-1}(Qm'(a_1,a_2)-m'(Qa_1,a_2)-(-1)^{|a_2|}m'(a_1,Qa_2)),
\nonumber\\
&& \{a_1,\mu(a_2,a_3)\}=\mu(\{a_1,a_2\},a_3)+(-1)^{(|a_1|-1)||a_2|}\mu(a_2,\{a_1, a_3\})\nonumber\\
&&\{\mu(a_1,a_2),a_3\}-\mu(a_1,\{a_2,a_3\})-(-1)^{(|a_3|-1)|a_2|}\mu(\{a_1,a_3\},a_2)=\nonumber\\
&&(-1)^{|a_1|+|a_2|-1}(Qn'(a_1,a_2,a_3)-n'(Qa_1,a_2,a_3)-\nonumber\\
&&(-1)^{|a_1|}n'(a_1,Qa_2,a_3)-(-1)^{|a_1|+|a_2|}n'(a_1,a_2,Qa_3),\nonumber\\
&&\{\{a_1,a_2\},a_3\}-\{a_1,\{a_2,a_3\}\}+(-1)^{(|a_1|-1)(|a_2|-1)}\{a_1,\{a_2,a_3\}\}=0.
\end{eqnarray}
The "homotopies" $m',n'$ are constructed by means of $\mu$, and $n$. In this paper we need no explicit expressions for them. \\
%In fact, they algebra above is a homotopy $BV$ algebra, i.e. the following relation holds: 
%\begin{eqnarray}
%(-1)^{|a_1|}\{a_1,a_2\}=b_0\mu(a_1,a_2)-\mu(b_0 a_1, a_2)-(-1)^{|a_1|}\mu(a_1,a_2)
%\end{eqnarray}

\noindent{\bf 2.2. Generalization of the LZ construction and deformed BRST charges.} 
Let us consider a CFT, which includes both chiral and antichiral parts, i.e. the space of states will be of the form ${\bf C}^*=C^*\otimes \b C^*$, where $\b C^*$ corresponds to the antichiral  part.  One can define the total BRST charge 
\begin{eqnarray}\label{Qtot}
\mathcal{Q}=Q+\bar{Q}=\oint (dz (c(z)T(z)+:c\p c b(z):)-d\b z(\t c(\b z)\t T(\b z)+:\t c\b \p \t c \t b(\b z):)). 
\end{eqnarray}
One can generalize both operations \rf{mu} and \rf{gerst} in such a way that they will satisfy the Gerstenhaber algebra up to homotopy with respect to 
$\mathcal{Q}$. 

The expression for $\mu$ is as before: $\mu(A_1,A_2)=P_0 A_1(\epsilon)A_2$, where $\epsilon\notin \mathbb{R}$, and $P_0$ is a projection on 
$\epsilon$-independent part. In order to give an expression for a generalization of homotopy Gerstenhaber bracket we  introduce the following notation. Let us associate the following 0-, 1- and 2-forms corresponding to the state $A\in {\bf C}^*$: $A^{(0)}\equiv A$, $A^{(1)}=dzb_{-1}A-d\bar{z}\t {b}_{-1}A$, $A^{(2)}=dz\wedge d\b zb_{-1}\t b_{-1}A^{(0)}$, where $b_{-1}$ and $\t b_{-1}$ are the appropriate modes of the chiral and antichiral b-ghost field correspondingly. If A is the primary state of conformal weight $(1,1)$, then we have a hierarchy of descent equations: $QA^{(0)}=0$, $QA^{(1)}=dA^{(0)}$, $QA^{(2)}=dA^{(1)}$. 
The expression for the analogue of the Gerstenhaber bracket in the case of ${\bf C}^*$ is, therefore (see also \cite{lmz}, \cite{zeit2}, \cite{zeit3}):
\begin{eqnarray}\label{gerstgen}
\{A_1,A_2\}=\frac{(-1)^{|A_1|}}{2\pi i}P_0\int_{C_{\epsilon}}A^{(1)}(z)A_2,
\end{eqnarray}
where $C_{\epsilon}$ is a circle contour of radius $\epsilon$ around the origin, and $P_0$ is the projection on the $\epsilon$ independent term, if one represents the right hand side as a power series in $\epsilon$. Actually, one can write $A_i=\sum_{\alpha} a_i^{\alpha}\otimes \bar{a}_i^{\alpha}$ $(i=1,2)$ where $a_i^{\alpha}\in C^*$, $\bar{a}_i^{\alpha}\in \b C^*$. Then
\begin{eqnarray}
&&\{A_1,A_2\}=\sum_{\alpha, \beta}(-1)^{|a^{\beta}_2||\bar{a}_1^{\alpha}|}  
\{a_1^{\alpha},a_2^{\beta}\}\mu(\bar{a}_1^{\alpha},\bar{a}_2^{\beta})+\nonumber\\
&&(-1)^{|a^{\alpha}_1|+|a_2^{\beta}|+|a^{\beta}_2||\bar{a}_1^{\alpha}|} 
\mu(a_1^{\alpha},a_2^{\beta})\{\b a_1^{\alpha},\b a_2^{\beta}\}.
\end{eqnarray}
One of the applications of the introduced operation is as follows. Suppose one 
has a CFT with the space of states ${\bf C}^*$. Let us perturb this CFT by a primary field $A(z)$ of conformal weight $(1,1)$. This corresponds to the perturbation of an action of the form $\frac{1}{2\pi i}\int \phi^{(2)}$, where  
$\phi^{(2)}=dz\wedge d\bar z A(z)$. Then, on the classical level, the conserved 
charge in this theory should be of the form $Q+\frac{1}{2\pi i}\oint A^{(1)}$. On the quantum level, in order to define the action of the deformed charge on the 
states of original theory, one has to define precisely the action of the integrated operator-valued 1-form. A natural choice will be 
$\mathcal{Q}+\frac{1}{2\pi i}\int_{C_{\epsilon}}A^{(1)}$. Unfortunately, we are not able to eliminate $\epsilon$-dependence by letting $\epsilon\to 0$, because of possible singularities. So one has to regularize it somehow. A natural choice is to take a projection $P_0$, on $\epsilon$-independent term. Therefore, the classical version of the deformed current on the quantum level will act as follows:
\begin{eqnarray}\label{defcharge}
B\to \mathcal{Q} B +\{A,B\},
\end{eqnarray}
where $B\in {\bf C}^*$. We have to say that there might be further corrections to the deformed charge, involving the operations of higher order in $A$. We claim that those contributions should also be governed by the Lian-Zuckerman formalism. We discuss it in the last part of this paper. 
In the next section, we study simple perturbations, where the formula \rf{defcharge} is just enough.

\section{BRST, logarithmic CFTs of the string theory and their "infinite metric" limits}
\noindent{\bf 3.1. The closed string via $\beta$-$\gamma$ systems.}
Let us consider the CFT with the action 
\begin{eqnarray}
S_{\beta,\gamma}=\frac{1}{2\pi}\int d^2 z (\beta_{i}\bar{\p}\gamma^{i}+\beta_{\b i}\p \gamma^{\b i }),
\end{eqnarray}
where $i, \bar{i}=(1,...,D/2)$. 
Let us perturb it by means of the operator $\phi_g^{(2)}=\frac{1}{2\pi i}dz\wedge 
d\bar{z}g^{i\bar{j}}\beta_i\beta_{\bar{j}}$, where $g^{i\bar{j}}$ is some flat metric. The resulting action is 
\begin{eqnarray}
S^g_{\beta,\gamma}=\frac{1}{2\pi}\int d^2 z \beta_{i}\bar{\p}\gamma^{i}+\beta_{\b i}\p \gamma^{\b i }-g^{i\bar{j}}\beta_i\beta_{\bar{j}}).
\end{eqnarray}
After a simple gaussian integration over $\beta$-variables, and redenoting $\gamma$ as $X$, one obtains a theory with the action
\begin{eqnarray}\label{closed}
S_{closed}=\frac{1}{2\pi}\int d^2 z 
g_{i\bar{j}}\p {X}^{i}\bar{\p} {X}^{\bar{j}},
\end{eqnarray}
where $g_{i\bar{j}}$ is the inverse matrix for  $g^{i\bar{j}}$. Here we change the  notations from $\gamma^i$ to $X^i$, since the operator meaning of 
$\gamma$- and $X$-variables is different. 
According to the considerations of Section 2, the deformed BRST charge 
is of the form 
\begin{eqnarray}
\mathcal{Q}\cdot+\{A^g,\cdot\}, \quad {\rm where}  \quad A^g(z)=\b c c g^{i\b j},\beta_i\beta_{\b j}
\end{eqnarray}
and $\mathcal{Q}$ is given by the formula \rf{Qtot}, where 
$T=-:\beta_i\p\gamma^i:$ and $\t T=-:\beta_{\bar{i}}\b \p\gamma^{\bar{i}}$. 
One can rewrite this operator in the following way:
\begin{eqnarray}
\mathcal{Q}^{g}= \mathcal{Q}-\frac{1}{2\pi i}\oint \frac{dz}{z}g^{i\b j}
 c\beta_i(z)\beta_{0,\bar{j}}+\frac{1}{2\pi i}\oint \frac{d\b z}{\b z}g^{i\b j}
 \b c\beta_{0,i}\beta_{\bar{j}},
\end{eqnarray}
where $\beta_{0,i}$ and $\beta_{0,\b j}$ are the $0$-modes of the corresponding conformal fields.
One can check on the operator level that this operator coincides with the operator
$\mathcal {Q}_X$, which is the BRST operator of the theory  \rf{closed}, i.e. 
 $\mathcal {Q}_X$ is expressed as in \rf{Qtot}, where $T=-g_{i\bar j}\p X^i\p X^{\bar{j}}$ and $\t T=-g_{i\bar j}\b \p X^i\b \p X^{\bar{j}}$, 
if we make the identification:
\begin{eqnarray}
&&\p X^{\bar j}=\beta_ig^{i\bar{j}}, \quad \bp X^{i}=\beta_{\bar j}g^{i\bar{j}},\nonumber\\
&& \p X^i=\p \gamma^i+\frac{\beta_{0,\bar j}g^{i\bar{j}}}{z},\quad 
\bp X^{\b j}=\bp \gamma^{\b j}+\frac{\beta_{0,i}g^{i\bar{j}}}{\bar z}
\end{eqnarray}
Thus, in this case, the formula \rf{defcharge} for the deformation of the BRST charge is exact and does not need to be improved by further corrections involving polylinear operations.\\

\noindent {\bf 3.2. The open string via $\beta$-$\gamma$ systems.}
In the case of the open string moving in $D$ dimensions, we consider a $\beta$-$\gamma$ system of the following type. Let $p_{\mu}$ and $X^{\mu}$ $(\mu=1,...,D)$ be the fields of conformal dimension $1$ and $0$ correspondingly. Their operator product is:
\begin{eqnarray}
X^{\nu}(z)p_{\mu}(w)\sim \frac{\delta^{\nu}_{\mu}}{z-w}, 
\end{eqnarray}
corresponding to the action 
\begin{eqnarray}\label{actpX}
S_{p,X}=\frac{1}{2\pi}\int d^2 z p_{\mu}\bar{\p}X^{\mu}.
\end{eqnarray}
Now we introduce an operator valued 1-form $\phi^{(1)}=\frac{1}{2}(dz c(\bar z)\eta^{\mu\nu}p_{\mu}(z)p_{\mu}(\bar{z})-d\b z c(z)\eta^{\mu\nu}p_{\mu}(z)p_{\mu}(\bar{z}))$, where $\eta^{\mu\nu}$ is a constant  Minkowski or Euclidean metric. Let us construct the following operator, which is a deformation of the BRST charge: 
\begin{eqnarray}\label{defpx}
Q^{\eta}=Q_{X,p}+\frac{1}{2\pi i} P_0\int_{C_\epsilon}\phi^{(1)}_{\eta}, 
\end{eqnarray}
where $Q_{X,p}=\oint (-cp_{\mu}\p X^{\mu}+:bc\p c:)$ is a BRST operator associated with $X$-$p$ theory. Counting the $\epsilon$-powers, one can find that this operator is:
\begin{eqnarray}
Q^{\eta}=Q_{X,p}-\frac{1}{2\pi i}\oint \frac{dz}{z}\eta^{\mu\nu}cp_{\mu}(z)p_{0,\nu},
\end{eqnarray}
where $p_{0,\mu}$ is the $0$th mode of the $p_{\mu}(z)$. On the language of the LZ operations it can be expressed by:
\begin{eqnarray}
Q^{\eta}=Q_{X,p}+\eta^{\alpha\beta}\mu(a_{\alpha},\{a_{\beta} ,\cdot \}),
\end{eqnarray}
where $a_{\mu}=cp_{\mu}$ . From here, it is clear that if the central charge 
of $X$-$p$ theory is equal to 26, $Q^{\eta}$ is nilpotent. 
Now, we show the relation of the operator above to the open string theory in dimension $D$. As before, we just need to compare the modes of appropriate fields. Namely, 
\begin{eqnarray}
&&p_{\mu}(z)=\sum_n p_{n,\mu} z^{-n-1}, \quad X^{\nu}(z)=\sum_n X^{\nu}_{n} z^{-n},\nonumber\\
&&[X_m^{\mu},p_{n,\nu}]=\delta_{m,-n}\delta^{\mu}_{\nu}.
\end{eqnarray}
Introducing the operators $a^{\mu}_n\equiv (\sqrt{2})^{-1}(nX^{\mu}_n+\eta^{\mu\nu}p_{\nu,n})$, $\bar {a}^{\mu}_n\equiv (\sqrt{2})^{-1}(nX^{\mu}_n-\eta^{\mu\nu}p_{\nu,n})$, (such that $n\neq 0$) one obtains that they form two commuting Heisenberg algebras:
\begin{eqnarray}
[a^{\mu}_n,a^{\nu}_m]=n\delta_{n,-m}\eta^{\mu\nu}, \quad [\b {a}^{\mu}_n,\b{a}^{\nu}_m]=-n\delta_{n,-m}\eta^{\mu\nu}, \quad [a^{\mu}_m,\b a^{\nu}_n]=0.
\end{eqnarray}
Using this notation, the operator  $Q^{\eta}$ can be represented as follows:
\begin{eqnarray}
Q^{\eta}=\oint (-c\eta_{\nu\mu}\p{\bf X}^{\nu}\p {\bf X}^{\mu}+c\b T+
c\p cb),
\end{eqnarray}
where 
\begin{eqnarray}
{\bf X}^{\mu}(z,\b z)=X^{\mu}_0-\eta^{\mu\nu}p_{0,\nu}ln|z|^2-\frac{1}{i\sqrt{2}}\sum_{m,m\neq 0}\frac{a^{\mu}_m}{m}(z^{-m}+{\b z}^{-m})
\end{eqnarray}
is the conformal field describing the open string on a half-plane, 
and $\b T$ depends on $\b a_m$ only. 
Therefore, if $F$ is a map eliminating all states generated by $\bar {a}_m$, then 
we have the following expression: 
\begin{eqnarray}\label{cmplxmap}
F Q^{\eta}=Q_{\bf X}F,
\end{eqnarray}
which means that $F$ is a chain map between the BRST differentials $Q^{\eta}$ and 
$Q_{\bf X}$ 
\footnote{We note here, that since the central charges of the Virasoro algebras are different for the $X$-$p$ theory ($c=2D$) and the open string ($c=D$),  
the differentials $Q^{\eta}$ and $Q_{\bf X}$ 
cannot be nilpotent at the same time on the whole space of the BRST complex.  However, later we will reduce both differentials  to the certain subcomplexes, where both of them are nilpotent.}.
Let us consider the following action on the half-plane $\bf{H}^+$:
\begin{eqnarray}
S_{p,X}^{\eta}=\frac{1}{2\pi}\int_{\bf{H}^+} d^2 z (p_{\mu}\bar{\p}X^{\mu}+\bar{p}_{\mu}\p \bar{X}^{\mu} -\eta^{\mu\nu}p_{\mu}\bar{p}_{\nu})
\end{eqnarray}
such that the boundary conditions for the fields are: $p_{\mu}(t)=\b  p_{\mu}(t)$ and $X^{\nu}(t)=\b X^{\nu}(t)$, where $t\in \mathbb{R}$. The third term in the action is a perturbing term.  Hence, one obtains that 
the perturbed BRST operator \rf{defpx} corresponds to the BRST operator of the theory described by the action above.

As we see, the consideration of the semi-infinite complex with the differential 
$Q^{\eta}$ on the space of the $X$-$p$ model has one big advantage: we get rid 
of logarithms, i.e. $X$-$p$ theory is a vertex algebra, 
and make the open string on a half-plane pure chiral. The disadvantage lies in the fact that the space of states of the $X$-$p$ theory is twice bigger than that of the 
${\bf X}$ theory, and, thinking about physical states, there will be additional auxiliary modes.

\section{The Yang-Mills equations via the \\
Lian-Zucker\-man homotopy algebra}
{\bf 4.1. Light modes and the Maxwell equations.}  In subsection 3.2. we have considered the $X$-$p$ version of the open string and defined the deformed BRST operator 
\begin{eqnarray}\label{qeta}
Q^{\eta}=Q_{X,p}+\eta^{\alpha\beta}\mu(a_{\alpha}\{a_{\beta} ,\cdot \}),
\end{eqnarray}
where $a_{\mu}=cp_{\mu}$, on the semi-infinite complex. 
Let us consider a subcomplex of the semi-infinite  complex, which corresponds to the states of conformal dimension $0$, i.e. these are the states annihilated by the operator $L_0=[Q_{X,p},b_0]$. These states can be explicitly written down. They correspond to the operators in the chiral algebra of the following form:
\begin{eqnarray}\label{lm}
&&\rho_u=u(X), \quad \phi'_{\mA}=cA_{\mu}(X)\p X^{\mu}, \quad\phi''_{\mB}
=c:B^{\mu}(X)p_{\mu}:,\nonumber\\ 
&& \phi_a=\p c a(X), \quad \psi'_{\mV}=c\p c V_{\mu}(X)\p X^{\mu}, \quad  \psi''_{\mW}=c\p c :W^{\mu}(X)p_{\mu}:,\nonumber\\
&&\psi_b=c\p^2 cb(X), \quad \chi_v=c\p c\p^2 c v(X),
\end{eqnarray}
where $\mA,\mV$ are the elements of the cotangent bundle and $\mB,\mW$ are the elements of tangent bundle. 
When we write the dependence of $X^{\mu}$ in the {\it fields}, involved in the operators above, like $a(X)$ or $A_{\mu}(X)$, one can assume that they are formal power series in $X^{\mu}$: they are still the elements of the $X$-$p$ vertex operator algebra, as it was shown in \cite{malikov}. 
In the following, we will refer to the states \rf{lm} as {\it light modes}, and denote the corresponding BRST subcomplex as $C^*_{L_0}$.
Now let us consider the action of an operator $Q^{\eta}$ on the BRST subcomplex of light modes. It is clear, that $Q^{\eta}$ commutes with $L_0$, therefore the  subcomplex of light modes is invariant under its action. Moreover, both $Q_{X,p}$ and $Q^{\eta}$ are nilpotent on $C^*_{L_0}$. 
Let us  calculate the 1st cohomology group with respect to $Q^{\eta}$
on $C^*_{L_0}$. 
\begin{eqnarray}
&&Q^{\eta}(\phi'_{\mA}+\phi''_{\mB}-\phi_a)=\psi'_{\Delta\mA}+\psi''_{\Delta\mB}+\frac{1}{2}\psi_{\p_{\mu}B^{\mu}+\eta^{\mu\nu}\p_{\mu}A_{\nu}}-\psi_a-\psi'_{\ud a}
-\psi''_{\ud a^*},\nonumber\\
&&Q^{\eta}\rho_{u}=\phi'_{\ud u}
+\phi''_{{\ud u}^*}-\phi_{\Delta u},
\end{eqnarray}
where ${({\ud u}^*})^{\mu}=\eta^{\mu\nu}\p_{\nu}u$ and 
$\Delta\equiv \eta^{\mu\nu}\p_{\mu}\p_{\nu}$. Therefore, we have the following equations which determine a cycle:
\begin{eqnarray}
&&\Delta A_{\nu}-\frac{1}{2}\p _{\nu}(\p_{\mu}B^{\mu}+\eta^{\mu\lambda}\p_{\mu}A_{\lambda})=0\nonumber\\
&&\Delta B^{\mu}-\frac{1}{2}\eta^{\mu\lambda}\p_{\lambda}(\p_{\nu}B^{\nu}+
\eta^{\rho\nu}\p_{\rho}A_{\nu})=0\nonumber\\
&&a=\p_{\mu}B^{\mu}+\eta^{\mu\nu}\p_{\mu}A_{\nu},
\end{eqnarray}
which means that the 1-form $(A_{\mu}+\eta_{\mu\nu}B^{\nu})\ud X^{\mu}$ satisfies the Maxwell equations and the fields $\Phi^{\mu}=\eta^{\mu\nu}A_{\nu}-B^{\mu}$ are massless scalar fields. Therefore, $H^1(C^*_{L_0})$ consists of 1-forms satisfying the Maxwell equations modulo gauge transformations and $D$ massless scalar fields. However, from the considerations of subsection 3.2., one can deduce that the map $F$ eliminates those scalar fields: they correspond to the system of auxiliary modes we obtained when considered the open string in the $X$-$p$ form. 

Studying the action of $Q^{\eta}$ in detail, one can obtain, that the complex of 
light modes is isomorphic to the one, which splits into the following subcomplexes:
\[
\xymatrixcolsep{30pt}
\xymatrixrowsep{-5pt}
\xymatrix{
0\ar[r]&\Omega^0 \ar[r]^{\ud} & \Omega^1 \ar[r]^{*\ud*\ud}  &\Omega^1 \ar[r]^{*\ud*} & \Omega^0\ar[r]&0\\
 &\quad &     & \quad    &&\\
 && \bigoplus & \bigoplus     &&\\
 &&           & \quad    &&\\
 &0\ar[r]& \oplus^D_{i=1}\Omega_i^0 \ar[r]^{\Delta}   & \oplus^D_{i=1}\Omega_i^0\ar[r]&0\\
&& \bigoplus & \bigoplus     &&\\
&0\ar[r]& \Omega^0 \ar[r]^{id}   & \Omega^0\ar[r]&0
}
\]
$\Omega^i$ denotes the $i$-forms on the $D$-dimensional space, 
where the symbol $*$ means a Hodge star with respect to the metric $\eta$. 
Evidently, after the action of the map $F$ only the upper complex 
survives. This upper complex is known as the $detour$ complex and was a starting point to build the $A_{\infty}/L_{\infty}$ structure of the gauge theory \cite{bvym}, \cite{cftym}. 
\\

\noindent{\bf 4.2. Deformation of the LZ structure and the Maurer-Cartan equations.}
From the explicit expressions of $\mu$-, $m$-, and $n$ -operations, one can see 
that they leave the complex $C^*_{L_0}$ invariant. So, from now on we restrict the LZ operations to light modes. 
From the previous subsection we already know that the deformed BRST operator 
$Q^{\eta}$ gives rise to the Maxwell equations. We want to find out whether it is possible to deduce the Yang-Mills equations via the Lian-Zuckerman construction.  
Since the BRST differential is deformed according to the formula \rf{qeta}, the Lian-Zuckerman operations are not homotopy associative with respect to $Q^{\eta}$. Therefore, one has to redefine (deform) them appropriately, in order to satisfy the relations \rf{lzrel} with $Q=Q^{\eta}$. 

Let us illustrate such deformation on a simple example. The deformed differential can be expressed in the following way: $Q^{\eta}=Q+R$, where we made a shorthand notation $Q_{X,p}\equiv Q$. 
The operation $R$ is not a derivation for the operation $\mu$. However, it is a derivation up to $Q$-homotopy, since $R=\eta^{\alpha\beta}\mu(cp_{\alpha},\{cp_{\beta},\cdot\})$ and 
$\{cp_{\mu},\cdot\}$ is just an action of the $p_{0,\mu}$-mode:
\begin{eqnarray}
&& R\mu(a_1,a_2)-\mu(Ra_1,a_2)-(-1)^{|a_1|}\mu(a_1,Ra_2)=\nonumber\\
&& Q\nu^{\eta}(a_1,a_2)-\nu^{\eta}(Qa_1,a_2)-(-1)^{|a_1|}\nu^{\eta}(a_1,Q a_2),
\end{eqnarray}
where $\nu^{\eta}$ is given by:
\begin{eqnarray}
\nu^{\eta}(a_1,a_2)=\eta^{\alpha\beta}(n(cp_{\alpha},p_{0,\beta}a_1,a_2)-\mu(m(cp_{\alpha}, a_1),p_{0,\beta}a_2))).
\end{eqnarray}
One can show that the bilinear operation $\mu$  should receive 
the following correction (of the first order in $\eta$) $\mu\to \mu^\eta=\mu-\nu^{\eta}$. In  this section, we will be interested in the quasiclassical limit of the Maurer-Cartan equation, associated with the $\eta$-deformed Lian-Zuckerman homotopy algebra. 
It means that one should introduce a parameter $h$ in the $X$-$p$ operator product: 
 \begin{eqnarray}
X^{\nu}(z)p_{\mu}(w)\sim \frac{h\delta^{\nu}_{\mu}}{z-w}
\end{eqnarray}
and to change the operator $Q^{\eta}\to Q_h^{\eta}=Q^h_{X,p}+h^{-1}R$, where 
$Q^h_{X,p}$ is the BRST differential with the rescaled Virasoro element: $T=-h^{-1}:p_{\mu}\p X^{\mu}:$. It can be shown that the corrections of higher order in $\eta^{\alpha\beta}$ lead to higher order powers of $h$. In the following, we will be interested in the leading order of powers of $h$, i.e. the smallest power with nonzero coefficient.

The Maurer-Cartan equation for the deformed LZ homotopy algebra is of the following form:
\begin{eqnarray}\label{mc}
Q_h^{\eta}\Psi+\mu_2^{\eta}(\Psi,\Psi)+\mu_3^{\eta}(\Psi,\Psi, \Psi)+...=0, 
\end{eqnarray}
where $\Phi\in C^1_{L_0}[h]\otimes U({\mathfrak{g})}$ ($\mathfrak{g}$ is some Lie algebra) and $\mu^{\eta}_2\equiv \mu^{\eta}$, $\mu^{\eta}_3\equiv n^{\eta}$ are 
the $\eta$-deformed operations $\mu$, $n$. Here dots stand for higher order operations in $\eta$-deformed Lian-Zuckerman homotopy associative algebra. 

By the quasiclassical limit of the Maurer-Cartan equation we mean the equation 
\rf{mc} considered on the factorcomplex $C^*_{L_0}[h]/h^2C^*_{L_0}[h]\otimes U(\mathfrak{g})$. 

So, we are interested in the terms of the first order in $h$ in the 
left hand side of \rf{mc}. 
One can show that higher order polylinear operations on 
$C^1_{L_0}[h]\otimes U(\mg)$ in the quasiclassical limit are equal to zero. Moreover, the higher order 
$\eta$-corrections to the operations $m$ and $n$ involve higher derivatives and higher $h$-powers. One can calculate their leading order in $h$ by the simple analysis of conformal dimensions (which should remain to be 0) and possible contractions between the $p$- and $X$-variables. 

Therefore, the equation \rf{mc} simplifies as follows:
\begin{eqnarray}\label{mcsemi}
Q_h^{\eta}\Psi+\mu^{\eta}(\Psi,\Psi)+n(\Psi,\Psi, \Psi)=0, 
\end{eqnarray}
where $\mu^{\eta}=\mu-h^{-1}\nu^{\eta}$. Now, we explicitly show, that this equation for the generic value of $\Phi$ gives rise to the Yang-Mills equations in $D$-dimensions coupled to $D$ scalar fields. 

First, we expand $\Psi=\phi'_{\mA}+\phi''_{\mB}-\phi_a$. 
Then let us explicitly write term by term.
\begin{eqnarray}
&&Q_h^{\eta}(\phi'_{\mA}+\phi''_{\mB}-\phi_a)=h\psi'_{\Delta\mA}+h\psi''_{\Delta\mB}+\frac{h}{2}\psi_{\p_{\mu}B^{\mu}+\eta^{\mu\nu}\p_{\mu}A_{\nu}}\nonumber\\
&&-\psi_a-\psi'_{\ud a}
-\psi''_{\ud a^*}.
\end{eqnarray}
Since $[X_0^{\alpha},p_{0,\beta}]=h\delta^{\alpha}_{\beta}$, one can replace 
$p_{0,\beta}$ by $-h\p_{\beta}\equiv -h\frac{\p}{\p X_0^{\beta}}$. Hence: 
\begin{eqnarray}
&&\mu^{\eta}(\Psi,\Psi)=\mu(\Psi, \Psi)+\eta^{\alpha\beta}(n(cp_{\alpha},\p_\beta \Psi,\Psi)-\mu(m(cp_{\alpha}, \Psi),\p_\beta \Psi)))=\nonumber\\
&& 
h(\psi'_{[\eta^{\alpha\beta}A_{\beta}+B^{\alpha},\p_{\alpha}\mA]}+
\psi''_{[\eta^{\alpha\beta}A_{\beta}+B^{\alpha},\p_{\alpha}\mB]}+\psi'_{\{\mA,\mB\}_1}+\psi'_{\{\mA,\mB\}^*_1})\nonumber\\
&&+\psi'_{[\mA,a]}+\psi''_{[\mB,a]}+\frac{h}{2}\psi_{A_{\mu}B^{\mu}+B^{\mu}A_{\mu}},
\end{eqnarray}
where $\{\mA,\mB\}_{1,\alpha}=\eta^{\beta\gamma}\p_{\alpha}A_{\beta}B_{\gamma}$ and $\{\mA,\mB\}^{*\alpha}_1=\eta^{\alpha\beta}\{\mA,\mB\}_{1,\beta}$.
The operation $n$ gives the following expression:
\begin{eqnarray}
&& n(\Psi,\Psi,\Psi)=\psi'_{\{\mA,\mB\}_2}+\psi''_{\{\mA,\mB\}_2^*},\quad {\rm where}
\nonumber \\
&&\{\mA,\mB\}_{2,\alpha}=\eta^{\beta\gamma}(A_{\alpha}(A_{\beta}B_{\gamma}+B_{\gamma}A_{\beta})+A_{\beta}A_{\alpha}B_{\gamma}+B_{\gamma}A_{\alpha}A_{\beta}),\nonumber\\
&&\{\mA,\mB\}^*_{2,\alpha}=\eta^{\beta\gamma}(B^{\alpha}(A_{\beta}B_{\gamma}+B_{\gamma}A_{\beta})+A_{\beta}B^{\alpha}B_{\gamma}+B_{\gamma}B^{\alpha}A_{\beta}).
\end{eqnarray}
The first equation we can derive from here, shows that $a$ can be expressed 
in the terms of $A$- and $B$- fields:
\begin{eqnarray}
a=\frac{h}{2}({\p_{\mu}B^{\mu}+\eta^{\mu\nu}\p_{\mu}A_{\nu}}-A_{\mu}B^{\mu}-B^{\mu}A_{\mu}).
\end{eqnarray}
One can see, that we eliminate auxiliary modes, provided by the map $F$ which means that $B^{\mu}=\eta^{\mu\nu}A_{\nu}$. In this case, it can be seen that the equation \rf{mcsemi} reduce to the Yang-Mills equations:
\begin{eqnarray}
&&\Delta A_{\mu}-\eta^{\alpha\beta}\p_{\mu}\p_{\alpha}A_{\beta}+
\eta^{\alpha\beta}\p_{\alpha}[A_{\beta},A_{\mu}]+\nonumber\\
&&\eta^{\alpha\beta}
[A_{\alpha},\p_{\beta}A_{\mu}-\p_{\mu}A_{\beta}]+\eta^{\alpha\beta}
[A_{\alpha},[A_{\beta},A_{\mu}]]=0.
\end{eqnarray}
In the general case, when $B^{\mu}\neq\eta^{\mu\nu}A_{\nu}$, we have two types 
of fields, $\mathcal{A}_{\mu}=A_{\mu}+\eta_{\mu\nu}B^{\nu}$ and 
$\Phi^{\mu}=\eta^{\mu\nu}A_{\nu}-B^{\mu}$. One can show, that the equation  \rf{mcsemi} leads then to the dimensionally reduced YM equations from $2D$ to $D$  such that $\mathcal{A}_{\mu}$ and $\Phi^{\nu}$ $(\nu=1,...,D)$, are 
the resulting gauge field and scalar fields. 
Finally, we note, that the gauge symmetries $\mathcal{A}_{\mu}\to \mathcal{A}_{\mu}+\epsilon(\p_{\mu}u+[\mathcal{A}_{\mu}, u])$, $\Phi^{\mu}\to \Phi^{\mu}$, where $\epsilon$ is infinitesimal, can be also reproduced by means 
of the symmetries of the Maurer-Cartan equation:
\begin{eqnarray}
\Psi \to \Psi+\epsilon(Q^\eta_h u +\mu^{\eta}(\Psi,u)-\mu^{\eta}(u,\Psi)).
\end{eqnarray}
\section{Further development}

\noindent{\bf 5.1. An algebraic approach to beta-fun\-ctions of sigma-models}. 
From what we have seen so far, the LZ homotopy algebras reproduce the classical field equations at the semi-classical level. What if we go further, and take into account all the $h$-corrections (or the $\alpha'$-corrections in the string-theoretic language)? 
The Maurer-Cartan equation we obtain, has the physical meaning of the conservation law of the deformed BRST charge, according to the considerations 
of Section 2 (see also \cite{zeit3}). Thus, one can treat the corresponding Maurer-Cartan equations as the equations of conformal invariance\footnote{This is close 
to the considerations of A.Sen (see e.g. \cite{sen}) in the context of the relation of the SFT and the conditions of conformal invariance.}.  
Therefore, the approach we used here gives us a possibility to study beta-functions for the perturbed  conformal field theories (in particular, sigma-models) algebraically. Actually, our considerations show that in the open string case (when metric is flat) the beta function vanishing condition has the form
\begin{eqnarray}
Q^{\eta}\Phi+\mu^{\eta}(\Phi,\Phi)+...=0,
\end{eqnarray}
where $\Phi$ is of ghost number $1$ and dots stand for higher polylinear operations. Whereas in the closed string case 
the condition of conformal invariance should be (see also \cite{zeit2}, \cite{zeit3}):
\begin{eqnarray}
Q\Psi+\frac{1}{2}\{\Psi,\Psi\}+...=0,
\end{eqnarray}
where $\Psi$ is is of ghost number $2$, 
$\{\cdot,\cdot\}$ is the operation \rf{gerstgen} and dots stand for higher 
polylinear operations. We will return to these equations in \cite{zeithomotopy}.\\

\noindent{\bf 5.2. Strings in the curved and the singular backgrounds.} 
A great advantage of the string theory reformulated in the first order formalism we introduced, is that we can at once isolate all light modes and work 
only with them. It does not hold in standard approach to the string models, since there is no natural operator which can serve such a projector. For example, this happens in the string field theory \cite{witten}, where massive modes couple to light modes and the only way to deduce, say, the field equations for light modes is to integrate out the massive ones, that is extremely hard to do (see e.g. \cite{taylor}, \cite{berkovits}, where this was done numerically). 
  
The algebraic approach we considered has another useful feature: it can be also applied when the corresponding metric is not flat. For the open string first order model we studied in Section 3, the natural possibility is to consider a  perturbation by means of the 2-form associated with the 
field of conformal dimension $(1,1)$ of the following kind:
\begin{eqnarray}
\phi^{(2)}_{e,\eta}=\frac{1}{2\pi i}d z\wedge d\b z\eta^{ab}:e_a^{\mu}(X)e_b^{\nu}(\b X)p_{\mu}\b p_{\nu}:(z),
\end{eqnarray}
i.e. representing the metric by means of orthonormal frames. The associated metric can even contain a singularity: since in our formulas we use the inverse metric, we might get rid of it. \\

\noindent{\bf 5.3. The YM $C_{\infty}$ algebra and the Courant algebroid.} 
One of the aims of this article was to (partly) explain the mystery of the  appearance of the YM $C_{\infty}$ algebra on the quasiclassical level \cite{cftym}, in the original logarithmic open string theory. In this paper we partly solved this problem, by getting rid of logarithms. We limited our calculations just to demonstrate, that the Maurer-Cartan equation leads to the YM equations, but in principle, it is not hard to show (the same way we did it in \cite{cftym}) that our deformed operations $\mu^{\eta}$ and $n$ reproduce, on quasiclassical level, the whole $C_{\infty}$ algebra.   

It appears, however, that our approach leads to another interesting observation. Let us return back to the complex $\mathcal{C}^*_{L_0}=C^*_{L_0}[h]/h^2C^*_{L_0}[h]$ and let's have a look on the homotopy Gerstenhaber structure, applied to the elements of $\mathcal{C}^*_{L_0}$:
\begin{eqnarray}\label{dorf}
\{\phi'_{\mA}+\phi''_{\mB}, \phi'_{\b \mA}+\phi''_{\b \mB}\}=h(
\phi'_{L_{\mB}\b \mA-\ud i_{\b \mB}\mA}+\phi''_{[\mB,\b \mB]_{Lie}}),
\end{eqnarray}
where $[\mB,\b \mB]_{Lie}$ is just a Lie bracket of the corresponding vector fields, 
and $L_{B}$ is a Lie derivative. The expression \rf{dorf} precisely coincides with the Dorfman bracket. So, one of the immediate consequences of the $\beta$-$\gamma$ approach is  the nontrivial relation between the Courant-Dorfman algebroid and the YM $C_{\infty}$  algebra, which will be studied further in \cite{courantzeit}.\\

\section*{Acknowledgements}

I am indebted to I.B. Frenkel, M.M. Kapranov and G.J. Zuckerman for fruitful discussions.

\end{document}